\documentstyle[12pt]{article}
\begin{document}
\topmargin 0pt
\oddsidemargin 5mm
\setcounter{page}{1}
\begin{center}
{\Large  Multiscaling at ferromagnetic-spin glass transition point of Random
 Energy Model and  complexity}\\
{D.B. Saakian}\\
{Yerevan Physics Institute,\\
Alikhanian Brothers St. 2, Yerevan 375036, Armenia.}
\end{center}
\begin{abstract}

We calculate  moments of free energy's finite size correction for the 
transition point between ferromagnetic and spin glass phases. We find, that 
those moments scale with the number of spins with different critical indices,
characteristic for the multiscaling. This critical point  corresponds to 
threshold of errorless coding for a gaussian noisy channel. We are give 
the definition of statistical complexity using this free energy approach.
\end{abstract}

The definition of statistical complexity is serious open problem in statistical mechanics [1]. There have been some attempts, based mainly on entropy concepts [2-4]. \\ 
In [5] I analyzed some complicated theories of modern physics-strings,2d quantum disorder. 
They have some statistical mechanical structures, equivalent to Random Energy Model (REM) [6], so we can define 
REM class of complex phenomena. I distinguished 2 classes of complex phenomena: like to spin glass (SG) phase of REM [6] and the transition region between spin glass (SG) and ferromagnetic (FM) phases.  In this work I am analyzing the second one 
(one is using term complexity today just for it) and give the definition (criteria) for statistical complexity. This is the phase of  complex adaptive systems (CAS) [7]). 
What we know nowadays, such 
complex structures arise between order and disorder 
(edge of chaos according to Kauffman [8]).In the REM case this point corresponds to errorless decoding treshold in information theory [9].\\
Let us solve the thermodynamics of REM at this region including finite size corrections, later try to generalize our results 
to define complexity.\\
One has , that  N  spins $s_i=\pm 1$ interact trough 
$C_N^P\equiv\frac{N!}{P!(N-P)!},P\to \infty$ couplings with the hamiltonian [6]
\begin{equation}
\label{e1}
H=-\sum_{1\le i_1..\le i_p\le N}[j^0_{i_1..i_p}+ j_{i_1..i_p}]s_{i_1}..s_{i_p}
\end{equation}
Here $ j^0_{i_1..i_p}$ are ferromagnetic couplings.
\begin{equation}
\label{e2}
j^0_{i_1..i_p}=\eta_{i_1}...\eta_{i_p}\frac{J_0N}{C^P_N}, \quad \eta_i=\pm 1
\end{equation}
 and for quenched noisy couplings $j_{i_1..i_p}$ we have distribution
\begin{equation}
\label{e3}
P_2(j)=\frac{1}{\sqrt{\pi N}}\sqrt{\frac{C^P_N}{N}} \exp\{-j^2\frac{C^P_N}{N}\}
\end{equation}
It is easy to show, that for large values of p (best case-$P=N/2$) there is a factorization for energy level 
distribution. For every $\alpha\ne\beta$
\begin{equation}
\label{e4}
P(E_{\alpha},E_{\beta})=P(E_{\alpha})P(E_{\beta})
\end{equation}
 For the 1-st configuration with $s_i=1$ we have  a distribution [9]
\begin{equation}
\label{e5}
P_1(E_1)=\frac{1}{\sqrt{\pi N}}\exp[-(E_1+J_0N)^2/N]
\end{equation}
and for other $2^N-1$ levels
\begin{equation}
\label{e6}
P(E)=\frac{1}{\sqrt{\pi N}}\exp(-E^2/N)
\end{equation}
While calculating free energy we performing averaging via energy level distribution (instead of random couplings 
in usual case of disordered systems):
\begin{equation}
\label{e7}
<\ln Z>\equiv <\ln\sum_{\alpha}\exp{(-\beta E_{\alpha})}>_E
\end{equation}
here $\beta$ is inverse temperature.
It is possible found [5], that at high enough values of $J_0$ at low temperature system is in ferromagnetic phase 
with magnetization
\begin{equation}
\label{e8}
m_i=\eta_i
\end{equation}
Later for convenient we take $\eta_i$=1.\\
 Using the trick [6]
\begin{equation}
\label{e9}
\ln Z =\Gamma'(1)+\int_{-\infty}^{\infty}\ln t \frac{d \exp[-tz]}{d t}{\it d}t
\end{equation}
one can factorize integration via different energy levels $E_{\alpha}$
it is possible to derive
\begin{equation}
\label{e10}
<\ln Z>=\Gamma'(1)+\int_{-\infty}^{\infty}u\frac{d \Psi(u)}{d u}{\it d}u
\end{equation}
\begin{equation}
\label{e11}
\Psi(u)=[f(u+u_f)f(u)^M]
\end{equation}
where $u=\ln t, u_f=J_0N\beta, \lambda=B\sqrt{N},M=2^N-1$ and
\begin{equation}
\label{e12}
f(u)\equiv \frac{1}{\sqrt{\pi} }\int_{-\infty}^{\infty}
\exp[-y^2-e^u \exp(-\lambda y)]dy
\end{equation}
$$ =\frac{1}{2\pi i}\int_{-i\infty}^{i\infty}\Gamma(x)
\exp[-ux+\lambda^2x^2/4]dx$$
In the last integral the loop overpasses point $0$ from right.
Function $f(u)$ is monotonic, like step. With exponential accuracy it equals $1$
below $0$, then become $0$ above it (with the same accuracy). We need in four
asymptotic regimes
\begin{equation}
\label{e13}
f(u)\approx\frac{1}{\sqrt{\pi}\lambda}\Gamma (2u/\lambda^2)\exp(-u^2/\lambda^2),
 \lambda\ll u 
 \end{equation}
$$f(u)\approx\frac{1}{\sqrt{\pi}}\int_{u/\lambda}^{\infty}dx\exp(-x^2),
 \lambda\ll \mid u \mid \ll \lambda^2 $$
$$f(u)\approx 1-\frac{1}{\sqrt{\pi}\lambda}\Gamma (-2u/\lambda^2)\exp(-u^2/\lambda^2),
-\lambda^2/2<u\ll-\lambda$$
$$f(u)\approx 1-exp(u+\lambda^2/4),
-\lambda^2<u<-\lambda^2/2$$
As the $f(u+u_f)f(u)^M$ is like to step function, its derivative is like to $\delta$
function with center coordinate at some $-u_0$. The vicinity of that point
gives the main contribution to the integral in (11) (bulk value is equal to $u_0$).
Ferromagnetic phase appears, when the $-u_f$ (the wall of function $f(u+u_f)$ is left, than
$-\sqrt{N}\ln 2 \lambda ($ the center of $f(u)^M$).
For the  boundary between FM and SG phases we have the line
\begin{equation}
\label{e14}
J_0=\sqrt{\ln 2},\qquad
\infty>B>\sqrt{\ln 2}
\end{equation}
We using the property, that if there is only 1-st level, $<\ln Z>=u_f\equiv J_0NB$ 
\begin{equation}
\label{e15}
\Gamma'(1)+\int_{-\infty}^{\infty}ud[f(u+u_f)]=u_f
\end{equation}
Using this equality, after simple transformations we derive
\begin{equation}
\label{e16}
<\ln Z>=\Gamma'(1)+\int_{-\infty}^{\infty}ud\Psi(u)
\end{equation}
$$=u_f-\int_{-\infty}^{\infty}ud\Psi_1(u)$$
$$\Psi_1(u)=f(u+u_f)[1-f(u)^M]du$$
Let us first consider the exact border of 2 phases,
\begin{equation}
\label{e17}
J_0=\sqrt{\ln 2}
\end{equation}
We have for $\Psi_1(u)$ a product of 2 monotonic functions, 
decreasing (one-to the left, another-
to the right) with the distance from the point $u=-u_f$.
Let us introduce $F(u)$
\begin{equation}
\label{e18}
F^{'}(u)=f(u+u_f)
\end{equation}
At $ \lambda\ll \mid u \mid \ll \lambda^2$ we derive
\begin{equation}
\label{19}
F(u-u_f)=\int_{0}^{u/\lambda}dx
[\frac{\lambda}{\sqrt{\pi}}\int_{x}^{\infty}\exp [-t^2]dt
\end{equation}
$$-\frac{C}{\sqrt{\pi}}\exp(-u^2/\lambda^2)]$$
After transformations $<\ln Z>$ goes to
\begin{equation}
\label{e20}
<\ln Z>=\Gamma'(1)+\int_{-\infty}^{\infty}udf(u+u_f)
-\int_{-\infty}^{\infty}ud\Psi_1(u)
\end{equation}$$=u_f+\int_{-\infty}^{\infty}\Psi_1(u)du$$
Let us perform integration by parts:
\begin{equation}
\label{e21}
<\ln Z>=u_f+\int_{-\infty}^{\infty}F^{'}(u)\Psi_2(u)du
\end{equation}
$$=u_f+F(\infty)\Psi_2(\infty)-F(-\infty)\Psi_2(-\infty)
-\int_{-\infty}^{\infty}F(u)\Psi^{'}_2(u)du$$
$$=u_f+[F(\infty)-F(-u_f)]
-F^{'}(-u_f)\int_{-\infty}^{\infty}(u+u_f)\Psi^{'}_2(u)du$$
Where $\Psi_2(u)=(1-f(u)^{M})$ and
we truncated expansion in degrees of $u+u_f$, because $\Psi^{'}_2(u)$ is
similar to  $\delta$ function near the $-u_f$. The main correction comes from the 
second part:
\begin{equation}
\label{e22}
<\ln Z>-u_f\approx 
\frac{\beta \sqrt{N}}{\sqrt{\pi}}\int_{0}^{\infty}dx\int_{x}^{\infty}\exp [-t^2]dt\sim N^{\frac{1}{2}}
\end{equation}
Let us consider now finite size correction for $<(\ln Z-u_f)^2>$
It is possible to derive 
\begin{equation}
\label{e23}
<(\ln Z)^2>=\Gamma'(1)^2+2\int_{-\infty}^{\infty}u\frac{d \Psi(u)}{d u}{\it d}u
\end{equation}
$$+\int_{-\infty}^{\infty}u_1u_2\frac{d^2 \Psi(\bar u)}{d u_1 d u_2}{\it d}u_1{\it d}u_2$$
where
\begin{equation}
\label{e24}
\bar u=\ln (e^{u_1}+e^{u_2})
\end{equation}
We have
\begin{equation}
\label{e25}
<(ln Z-u_f)^2>-<(ln Z-u_f)>^2
\end{equation}
$$=\int_{-\infty}^{\infty}u_1u_2\frac{d^2 \Psi_1(\bar u)}{d u_1 d u_2}{\it d}u_1{\it d}u_2$$
$$-2(u_f-\Gamma')\int_{-\infty}^{\infty}u\frac{d \Psi_1(u)}{d u}{\it d}u
-(\int_{-\infty}^{\infty}u\frac{d \Psi_1(u)}{d u}{\it d}u)^2$$
In this last expression the main part gives
\begin{equation}
\label{e26}
-2u_f\int_{-\infty}^{\infty}u\frac{d \Psi_1(u)}{d u}{\it d}u=
\end{equation}
$$\frac{2\beta^2J_0N \sqrt{N}}{\sqrt{\pi}}\int_{0}^{\infty}dx\int_{x}^{\infty}\exp [-t^2]dt\sim N^{\frac{3}{2}}$$
Comparing (22) and (26) we see multiscaling property. This is the main result of our work. In [12] has been observed
multicriticity at FM-PM transition point for the case of weak, ferromagnetic disorder.\\
Let us now consider small deviation from (12)
\begin{equation}
\label{e27}
J_0=\sqrt{\ln 2}+\frac{u_1}{\sqrt{N}}
\end{equation}
We can define the magnetization m using the formulas 
\begin{equation}
\label{e28}
\begin{array}{l}
m=<\frac{exp(-\beta E_0)}{exp(-\beta E_0)+\sum_i\exp(-\beta E_i)}>
\end{array}
\end{equation}
Using the identity $\frac{1}{Z}=\int_0^{\infty}d te^{-t Z}$ 
It is easy to derive 
\begin{equation}
\label{e29}
m=-\int_0^{\infty}d t\frac{d}{d t}f'(u+u_f)f(u)^M=1-\int_{-\infty}^{\infty}f_0(u+u_f)\frac{d}{d u}f^M(u)
\end{equation}
In the main approximation we have
\begin{equation}
\label{e29}
m=\frac{1}{\sqrt{\pi}}\int_{-u_1}^{\infty}d x\exp(-\frac{x^2}{2})
\end{equation}
Using the formula for magnetization we derive
\begin{equation}
\label{e30}
m_i=m\equiv\frac{d<\ln Z>}{\beta \sqrt{N}d u_1}  =
\frac{1}{\sqrt{\pi}}\int_{-u_1}^{\infty}\exp [-t^2]dt
\end{equation}
and for its differential
\begin{equation}
\label{e31}
\frac{d m_i}{d j_0}=\frac{1}{\sqrt{\pi}}\exp [-(u_1)^2]
\end{equation}
We see, that at $u_1=0$ this quantity is maximal. So in our region (exact border SG-FM) our the dependence of magnetization 
 from the external (ordered,manageable) parameter is maximal. Perhaps this (instability) property is the orign of edge of chaos's popularity in complex adaptive systems (CAS)[2]. Here one has some ordered external parameter to manage the system as well as
 some random parameters. There is some emergent (completely collective) property, like magnetization.
The interaction measure with environment for our system is some fitness function from that emergent property. So it is highly reasonable, that CAS drifts to the maximal instability point, where the dependence of this emergent property from the ordered par
ameter is maximal.\\
So we are suggesting as complexity:\\
A.There is an emergent property, which is maximally unstable under the change of ordered external parameters.\\
 One can apply this criteria  of complexity to 
both disordered (like SG) or ordered (like ferromagnetic Ising) models. We see, that it distinguishes critical regions. 
This criteria could be applied even for single and finite element systems, we need only in mapping $J_0\to m$.
If we want to distinguish SGF-FM transition point from PM-SG one, we need to increase our criteria by additional constraint.\\
B.We should have free energy for our system at different scale N, and some ensemble of such systems (may be at different space or time 
scale). At complexity phase  free energy's finite size corrections should have multiscaling.\\
In multiscaling phenomena, as it is known [10], it is possible to define free energy like quantities , proportional to logarithm of scale.
So we can give second (rather speculative) form of additional criteria instead of B:\\
C. There should be in our system usual free energy and second fine form of free energy, highly suppressed in magnitude compared with the 
first one.\\
This second formulation is more elegant (while speculative), perhaps it could work for a single system also.\\
So in contrary to other approaches we are using free energy instead of entropy. Why?\\
Free energy is manageable (via external parameter) part of energy. In the situation, when there are both thermal noise and noise in external parameters, it is reasonable to consider just this (manageable via temperature) amount of energy. We see also in o
ur toy model, that entropy has not multiscaling.\\
Let us remember the multiscaling [10-11]in details to clarify criteria C. One considers some geometrical objects in a box with macroscopic size L and microscopic cutoff a with corresponding probabilities $p_x$. Then it is possible to introduce 2 sets of i
ndices [10]:
\begin{equation}
\label{e32}
\begin{array}{l}
\tau(q)=\lim \frac{1}{\ln (a/L)}\ln (\sum_xp_x^q)
\end{array}
\end{equation}
If our probabilities $p_x$ scale as $\int {\it d}\alpha'\rho(\alpha')(\frac{a}{L})^{-f(\alpha')}$, then two sets of indices $\tau(q)$ and $f(\alpha)$ are connected by a Legendre transformation [10]
\begin{equation}
\label{e33}
\begin{array}{l}
\alpha=\frac{{\it d}\tau(q)}{{\it d}q}\\
f(\alpha)=\alpha q-\tau(q)
\end{array}
\end{equation}
One can take as $p_x$ free energy's finite size correction, then $\tau(q)$ is free energy density for the new system with $-\ln (a/L)$ volume
 and q is inverse temperature [10-11]. New free energy describes the change of critical indices under the change of 
 his internal parameter q. What is important, it is tiny, proportional to $\ln L$ instead of L for usual free energy F. This highly 
 resembles schemata (compressed crucial information) in CAS according to M. Gelmann [7], as well as free energy for strings and conformal
  models [5].\\
We give qualitative and descriptive definition of complexity on the ground of free energy. If the structure of theory is more or less
known, one can use our criteria to establish a complexity.\\
It will be very interesting to construct generalization of REM (GREM) with different scaling for free energy corrections, as well as try to find for any case of multiscaling corresponding REM (for finite size correction).  In [11] there is an example of s
uch connection. \\
If such correspondence could be found, there is a chance for self-referentiality (finite size corrections of REM could be described by means of another REM). In [13] the author argued, that  this property was crucial to make consciousness possible in mind
-brain systems, and suggested, that it could be in other complex systems.\\
So there is a good chance, that REM catchs the main features of complex adaptive systems. 

\end{document}